# An inverse free electron laser acceleration-driven Compton scattering X-ray source

I. Gadjev[1], N. Sudar[1], M. Babzien[2], J. Duris[1], P. Hoang[1], M. Fedurin[2], K. Kusche[2], R. Malone[2], P. Musumeci[1], M. Palmer[2], I. Pogorelsky[2], M. Polyanskiy[2], Y. Sakai[1], C. Swinson[2], O. Williams[1], J. B. Rosenzweig[1]

[1]UCLA Department of Physics and Astronomy, 405 Hilgard Ave., Los Angeles, CA 90095
[2]Brookhaven National Laboratory, Upton, NY 11973

## Abstract

The generation of X-rays and γ-rays based on synchrotron radiation from free electrons, emitted in magnet arrays such as undulators, forms the basis of much of modern X-ray science. This approach has the drawback of requiring very high energy, up to the multi-GeV-scale, electron beams, to obtain the required photon energy. Due to the limit in accelerating gradients in conventional particle accelerators, reaching high energy typically demands use of instruments exceeding 100's of meters in length. Compact, less costly, monochromatic X-ray sources based on very high field acceleration and very short period undulators, however, may revolutionize diverse advanced X-ray applications ranging from novel X-ray therapy techniques to active interrogation of sensitive materials, by making them accessible in cost and size. Such compactness may be obtained by an all-optical approach, which employs a laser-driven high gradient accelerator based on inverse free electron laser (IFEL), followed by a collision point for inverse Compton scattering (ICS), a scheme where a laser is used to provide undulator fields. We present an experimental proof-of-principle of this approach, where a TW-class $CO_2$ laser pulse is split in two, with half used to accelerate a high quality electron beam up to 84 MeV through the IFEL interaction, and the other half acts as an electromagnetic undulator to generate up to 13 keV X-rays via ICS. These results demonstrate the feasibility of this scheme, which can be joined with other techniques such as laser recirculation to yield very compact, high brilliance photon sources, extending from the keV to MeV scale. Furthermore, use of the IFEL acceleration with the ICS interaction produces a train of very high intensity X-ray pulses, thus also permitting a unique tool that can be phase-locked to a laser pulse in frontier pump-probe experimental scenarios.

The rapid progress in X-ray science over the last century, beginning with cathode ray tubes and arriving at 4[th] generation light sources – in the form of the X-ray free-electron laser, XFEL – has been driven by many breakthroughs in the fields of electron acceleration and synchrotron radiation generation. Dramatic advances in these techniques have fueled many discoveries across a wide swath of scientific disciplines ranging from physics[1,2], to chemistry[3], biology[4], and material science[5]. The main drawback of modern high brightness X-ray sources is their large size and cost, driven both by the size and complexity of the high energy particle accelerators and the elaborate undulator magnets used. In an attempt to create a more compact, high flux, high brilliance X-ray source, *5[th] generation X-ray sources* based on all-optical schemes involving laser-driven accelerators and laser-enabled undulators have been the subject of intensive study[6–11].

In such an all-optical X-ray light source, one may replace the undulator or wiggler magnets, which have cm-scale period, with the $\mu m$ wavelength oscillating electromagnetic field of a high power laser. This change allows one to reach similar photon energies as reached by magnetostatic undulators with two orders of magnitude lower energy electron beams. This electron-laser radiative interaction, termed inverse Compton scattering (ICS), is suitable for the production of very high-energy photons, when the scattering electrons are ultra-relativistic. For example, at the Brookhaven National Laboratory's Accelerator Test Facility (BNL ATF), site of the experiments reported here, tens of MeV-scale electrons can scatter far-infrared $CO_2$ laser photons to reach X-ray energies in the tens of keV [12,13]. With GeV-class electron beams, one may produce high brilliance, spectrally narrow γ-ray beams, with photon energies in the MeV range and beyond[14]. Given such possibilities, an ICS-based X-ray source possesses desirable characteristics for many X-ray applications that demand narrow bandwidth, short pulse, and directional high-flux X-ray beams.

While taking advantage of the ICS interaction greatly reduces the demands on the electron beam, in order to reach MeV-class photon energies and also improve the brilliance of the X-rays an electron beam in the several 100's of MeV is still needed. It is possible to use the same laser pulse utilized for the Compton interaction to enable a compact, high-gradient electron accelerator. The inverse free-electron laser (IFEL) scheme is indeed such an approach, which is attractive due to its ability to produce higher than state-of-the-art acceleration gradients in a material free interaction region. It thus permits use of the laser for acceleration without the usual limitations arising from nearby matter, *i.e.* wakefields, and material breakdown at high field.

The strong interest in the development of 5th generation light sources along with the rapid progress in recent years on the physics of both ICS and IFEL, and the recent demonstration at the BNL ATF of ICS enhanced by laser recirculation[15], makes it quite timely to examine the intersection of ICS and IFEL in a single experiment. In this paper, we show the use of an IFEL that produces a high quality, microbunched accelerated electron beam which in turn feeds a Compton scattering interaction point (IP) based on the same laser system, for the production of a narrow bandwidth, directional, pulsed X-rays. We demonstrate in particular, an all-optically driven, electron beam-based X-ray source. This is an important step forward in current research into advanced accelerators and their use in real world applications, in particular compact light sources. The encountered experimental challenges address frontier demands in accelerator physics concerning beam quality[16] and push forward experimental expertise that will be needed to arrive at yet higher energy applications of advanced accelerators.

## Background

X-rays generated by ICS are localized in angle to a $1/\gamma$ cone about the electron propagation direction – a directionality characteristic of radiation by relativistic charged particles. For ultra-relativistic electrons, $\gamma =$

$\frac{U_e}{m_e c^2} \gg 1$, and neglecting the recoil of the electron (the Thomson limit, which is valid when the laser photon's momentum in the electron rest frame is smaller than the electron's rest energy), the single-particle angular dependence of the X-ray spectrum is: $\frac{hk_{xray}}{hk_L} = \frac{4\gamma^2}{1+\gamma^2\theta^2}$. While there is an inherent off-axis redshift, the angular distribution is confined to within this $1/\gamma$ angle; these higher energy components thus also correspond to the highest flux density of X-rays.

In practice, the scattering takes place in the context of a highly focused, short pulse beam of electrons colliding with a laser pulse of similar spatio-temporal characteristics. There are a number of aspects of the interaction arising from the distribution of electron and photon angles in the beams, as well as the influence of the finite time of laser-electron interaction, that affect the flux, bandwidth, and divergence of the X-ray photon distribution generated[17]. The most basic of these considerations is that the total number of generated X-rays is proportional to the number of electrons and laser photons available for interaction as well as the cross-sectional overlap of the two beams, *i.e* the luminosity $L$ of the collision. If the transverse profiles of the electron bunch and laser pulse are well approximated by bi-gaussian distributions, the optimally efficient configuration is to have the electrons and laser have the same transverse size, $\sigma_L = \sigma_x$, which gives: $N_{xray} = \sigma_T N_e N_L / 4\pi\sigma_x^2 = \sigma_T L$, where $\sigma_T$ is the Thomson cross-section.

The IFEL is an efficient process for transferring energy from a laser pulse to a co-propagating relativistic electron beam[18]. In present radio-frequency accelerators, boundary conditions are employed that permit rotation of the electric field to have a significant longitudinal component, and thus energy can be exchanged with particles traveling in the longitudinal direction. Since the IFEL interaction occurs in vacuum, however, for energy transfer to occur the transverse electric field of the driving laser pulse is coupled to the electron motion via the transverse oscillation induced by a periodic magnetic field. In order to have continuous acceleration in an IFEL, a stationary phase of the electrons with respect to laser pulse must be maintained[19]. This is managed in practice by adjusting the period of the external magnetic field, a procedure termed undulator "tapering". Inside an undulator, the electrons with resonant energy, maintain a nearly constant interaction phase with co-propagating light waves, when the undulator condition is met: $\frac{k_u}{k_L} = \frac{1+K_u^2}{2\gamma^2}$. As the electron energy changes during acceleration to maintain resonance one should taper the undulator in such a way as to continuously preserve the accelerating phase of the pondermotive force. Inspection of the resonance condition indicates that one may change either the normalized undulator vector potential, $K_u$ or the undulator period, $\lambda_u = 2\pi/k_u$, or both, to accomplish this goal.

Although IFEL acceleration gradients are modest compared to an alternative laser-based acceleration concept, laser plasma acceleration (LPA) [20], the IFEL approach has a number of important advantages. First, the IFEL is a free space accelerator (*cf.* Figure 1), and as such no energy is lost to the media, and very high gradients can be sustained without a risk of material breakdown. With the ability to use weakly focused, very high energy laser pulses, a single IFEL stage can be made up to a few meters long, providing continuous interaction over such extended lengths. Further, absorption of energy from the electron beam is the only notable effect of the interaction that influences the radiation propagation. Therefore, for low electron beam currents, pump depletion effects can be minimized allowing the possibility to recycle the laser beam in a high repetition rate configuration[21,22]. This configuration enables greatly enhanced average radiation flux, which is an over-arching goal of the research described here. Finally, since the resonant characteristics of the IFEL interaction is (assuming a laser with modest intensity fluctuations) dominated by the static magnetic fields of the undulator, the accelerated beam longitudinal phase space can be well controlled through proper undulator design. This process can result in small energy spread and much reduced output energy fluctuations, especially compared to schemes such as LPA which translate laser intensity errors to electron output energy errors[16]. We note that the phenomenon of energy loss to wakefields has an analogy in the IFEL – synchrotron radiation losses. These effects limit the maximum practical beam energy in an IFEL to the 10's of GeV range, which is not a serious issue for light source applications.

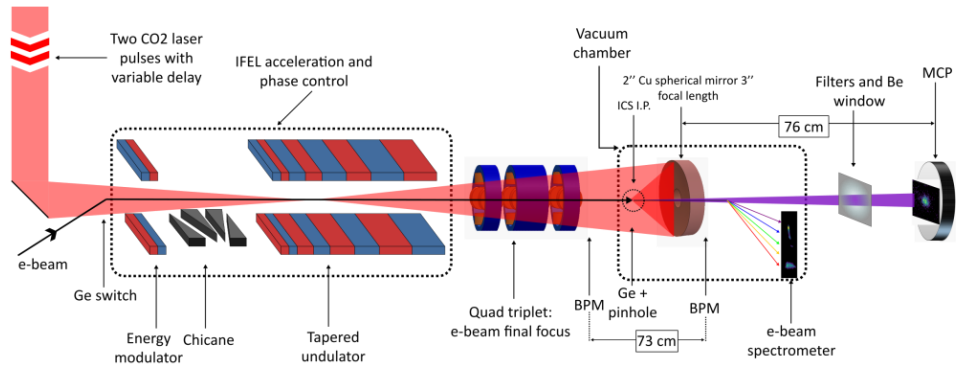

Figure 1. The schematic drawing of the second electron beam-line at the BNL-ATF. The two $CO_2$ laser pulses co-propagate with the electron beam and are reflected by the spherical Cu mirror. The electron beam is imaged onto the spectrometer after colliding with the leading $CO_2$ pulse. The generated X-rays are collected by an MCP.

## Results

This merging of the IFEL accelerator and the ICS X-ray IP to yield a unique source of X-ray photons was performed on a high brightness electron beamline at the BNL ATF. The general features of this beamline, which permits electron interactions with a high peak power $CO_2$ laser operating at $10.3$ μm, are shown in Figure 1. There are three points of laser-electron interaction in this experimental scenario: a short IFEL energy modulator is first encountered, which combined with a downstream chicane forms a pre-bunching system; a subsequent tapered IFEL undulator, known as the Rubicon undulator, where the laser provides pondermotive acceleration; and, finally, the ICS interaction point that yields X-ray production. This experimental design requires two co-linear laser pulses obtained from the same $CO_2$ laser amplifier. The pre-bunching and IFEL interaction sections are driven with the trailing $CO_2$ pulse. The leading $CO_2$ pulse, on the other hand, is reflected and re-focused onto the electron bunch by a spherical Cu mirror, $f/\#=1.5$, which is housed in a vacuum chamber $2.7$ m downstream of the undulator. The $\sim 0.5$ ns delay between the two laser pulses is nominally double the time of flight associated with the Cu mirror focal length ($f = 7.5$ cm). An electron beam activated Ge switch was inserted directly upstream of the pre-buncher, and the $CO_2$ transmission through the Ge screen was used to superimpose the electron beam and laser in time to $\sim 1$ ps [23,24]. Finer, sub-picosecond level synchronization was achieved based on optimization of the acceleration and the X-ray flux. Immediately after the ICS IP, the electron beam is deflected by a permanent magnet dipole and dumped onto a large energy acceptance spectrometer. The Compton X-rays are detected **76 cm** downstream using a micro-channel plate (MCP), which gives not only the total flux, but the angular distribution of these ICS-derived photons.

The design of the Rubicon undulator has been documented in previous publications by Duris et al. [18]. The undulator is a strongly tapered, helical, permanent magnet design that permits efficient, continuous accelerating and bunching interaction between the electrons and the co-propagating laser. It is the first of its type deployed in such experiments.

In order to increase the fraction of accelerated electrons captured in the Rubicon IFEL's longitudinal acceptance, we employed a pre-buncher assembly to shape the longitudinal phase space of the electron beam upon entrance into IFEL[25,26]. The pre-buncher consists of a single period permanent magnet planer undulator and a permanent magnet variable-gap chicane and is described by Sudar, *et al* [27]. The laser's electric field imparts periodic energy modulation on the beam as the two co-propagate inside the single period undulator. This energy modulation is then transformed to a density modulation, or microbunching through use of a chicane magnet system, that also controls the phase of the bunching relative to the phase of the electric field of the laser. Placing the peak of the density distribution at the appropriate pondermotive IFEL phase significantly enhances the capture rate. The results shown in Figure 2 demonstrate a clear increase in the fraction of electrons accelerated. The pre-bunched IFEL (b) accelerated **43%** of the beam as compared to only **17%** in the non-pre-bunched IFEL case (a).

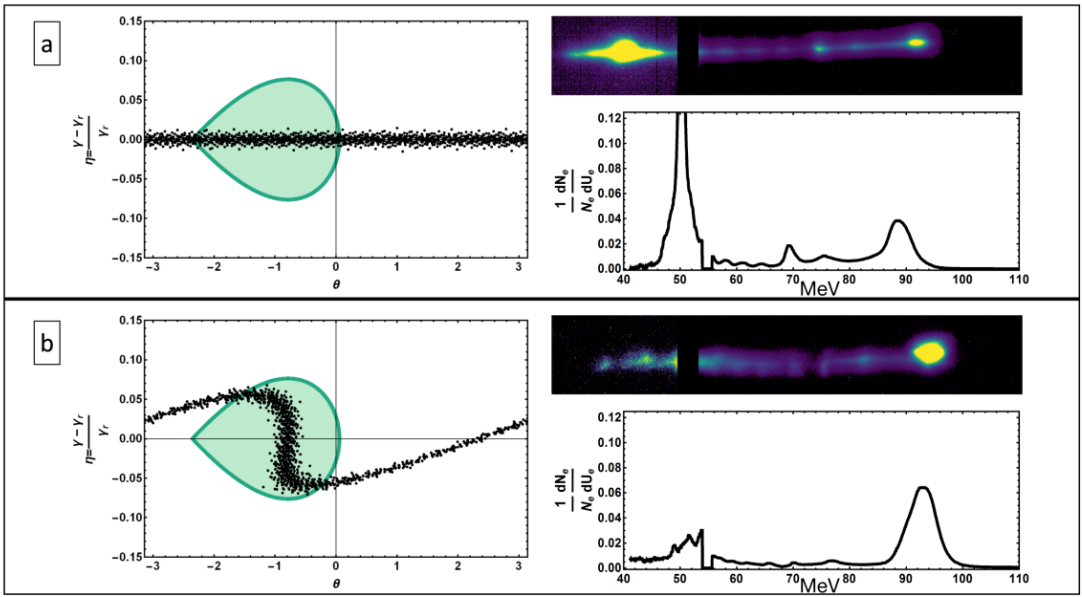

*Figure 2. The electron spectrometer image and lineout in b) show the significant increase in capture when the pre-bucnher is employed to load the initial electron beam into the $\pi/4$ phase of the pondermotive potential. The shaded area in the $\eta - \theta$ plots signifies the area of longitudinal phase-space that will be captured to full acceleration.*

Electron spectra and X-ray angular distributions from a sequence of six tests are displayed in Figure 3. Images from the electron energy spectrometer are shown in the bottom figures, with higher energy electrons appearing towards the top of each image. The electron beam was bunched and accelerated by the IFEL from 52 MeV to a maximum of 82 MeV with an rms 5% energy spread. The total energy gain was 30 MeV, which corresponds to an effective maximum accelerating gradient in the IFEL of 55 MeV/m for correctly phased electrons. The final spectrum is quite stable against laser fluctuations, as illustrated by the collection of shots shown in Figure 3; it is dictated, within certain bounds of intensity, by the undulator design. In displaying this spectrum, we note that the quadrupole triplet's focusing effects are adjusted to focus the more energetic and more rigid accelerated electron beam, and are too strong for the non-accelerated beam. This is shown in the spectrometer images, as the (mean energy) 82 MeV electrons are optimally focused while 52 MeV electrons have passed through a focus and appear diffuse at the spectrometer; they are also not optimally focused at the IP, and thus do not participate efficiently in the ICS interaction. The narrow band of captured and accelerated electrons are found to suffer negligible emittance growth, a feature of optimized IFEL systems that is essential for the effective focusing of electrons at the ICS IP. The preservation of e-beam emittance and its effect on the final focus into the ICS IP are describe in Figure *4*.

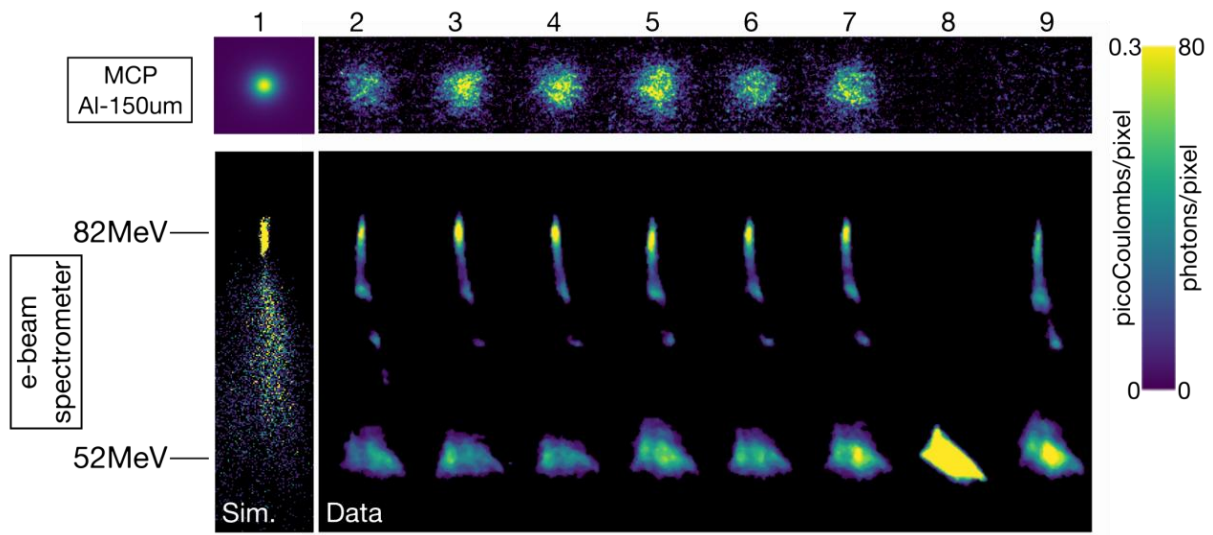

*Figure 3. Images of the electron spectrometer and the X-ray MCP. Higher energy electrons appear towards the top of the spectrometer images. The MCP images are taken downstream of a* 150 µm *thick Al attenuator. A sequence of six shots in which both IFEL acceleration and ICS X-rays were observed is presented in images 2-7. Image 8 shows an unaccelerated electron beam, which displays ICS X-ray flux completely attenuated by the Al filter. Image 9 demonstrates the X-ray background from an accelerated e-beam. The 3D simulation of the IFEL process and the expected X-ray signal in image 1 agrees with the data.*

The top images in Figure 3 show the spatial distribution of ICS X-rays deposited on the MCP after passing through a 150 µm thick Al foil. The Al foil acts as a strong attenuator for lower energy X-rays. Since the energy distribution of the Compton scattered X-rays reflects that of the electron beam (see Figure *4*), the portion of X-rays that passes through the Al foil corresponds to X-rays produced from the accelerated, 82 MeV electrons. In Figure 3, shots 2-7 indicate a clear ICS X-ray production above 10 keV from the accelerated electrons. When the IFEL acceleration is turned off, the Al attenuator blocks all ICS X-rays (image 8). On the other hand, if the ICS laser pulse is not present while IFEL acceleration is ON, it is seen that there is no significant X-ray background to the MCP (image 9). This spectral filtering thus demonstrates the successful operation of the IFEL driven ICS X-ray source. The measured X-ray flux extends over the range $10-13$ keV, with a central peak at $11.6$ keV, and contains $1.22(\pm 0.16)\times 10^6$ photons per shot, which is in good agreement with the simulation based estimate of $1.3\times 10^6$ photons per shot.

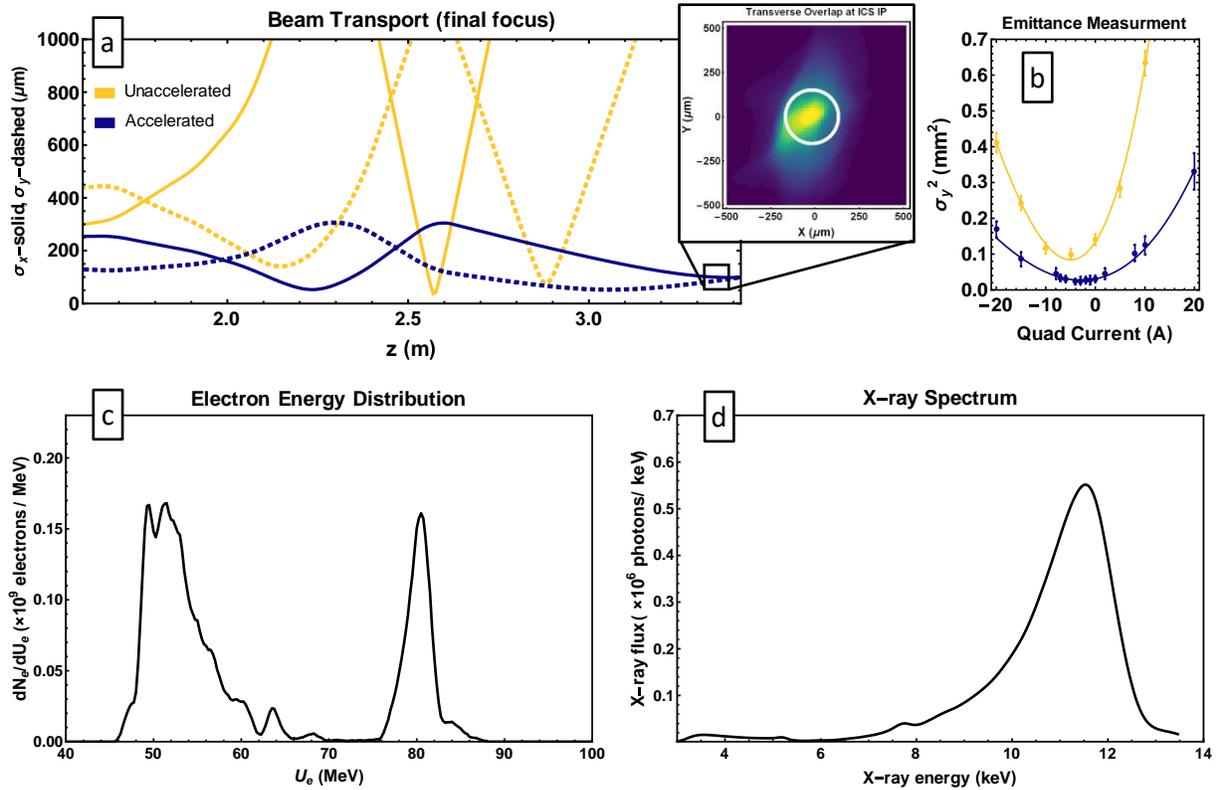

Figure 4. Although the electron energy spectrum shows a large portion of unaccelerated electrons c), the focus into the ICS IP is set to produce a transversely dense beam only for higher energy electrons a), which translates into an X-ray spectrum completely dominated by $12\ keV$ photons d). The inset of part a) is an image of the transverse profile of the electron beam at the ICS IP $\sigma_x\times\sigma_y=140\times 170\ (\pm 10)\ \mu m$, and the circle marks the estimated $CO_2$ FWHM there $\sim 150\ \mu m$. The emittance measurement b) confirms the preservation of the normalized emittance of the beam by the IFEL acceleration. Emittance measurement data was taken for the unaccelerated beam and accelerated beam, yielding normalized emittance values of $2.3\pm 0.1\ \mu m$ and $2.4\pm 0.15\ \mu m$ respectively.

As a further validation of both the electron beam energy increase and the concomitant increase in X-ray brightness obtained with the use of IFEL acceleration, we examine the angular distribution of the scattered photons. The ICS X-ray distribution exhibits a $\theta\sim 1/\gamma$ forward opening angle typical of the radiation by an accelerated relativistic particle. This property can be used to deduce the energy of the emitting X-rays by measuring their angular spread, assuming the angles in the electron beam at the IP are $<1/\gamma$. Figure 5 shows the distribution of X-rays on the MCP detector for two different energies: $5\ keV$ X-rays produced by the unaccelerated electrons at $52$ MeV without an attenuating foil; and $12$ keV X-rays produced by the fully accelerated electrons at $82$ MeV passing through a $150\ \mu m$ thick Al attenuating foil. The lineouts through the center of these transverse profiles clearly show the narrowing of the angular spread for higher energy X-rays. The angle at which 90% of the flux is encompassed for the $5$ keV X-rays occurs at $\theta_{52}=4.7\pm 0.1$ mrad, while for the $12$ keV flux this angle narrows to $\theta_{82}=3.4\pm 0.1$ mrad. This observation validates the increased energy in the electron beam, and also illustrates that this beam retains its small angles after IFEL acceleration and transport to the IP — a limit on beam emittance growth is set.

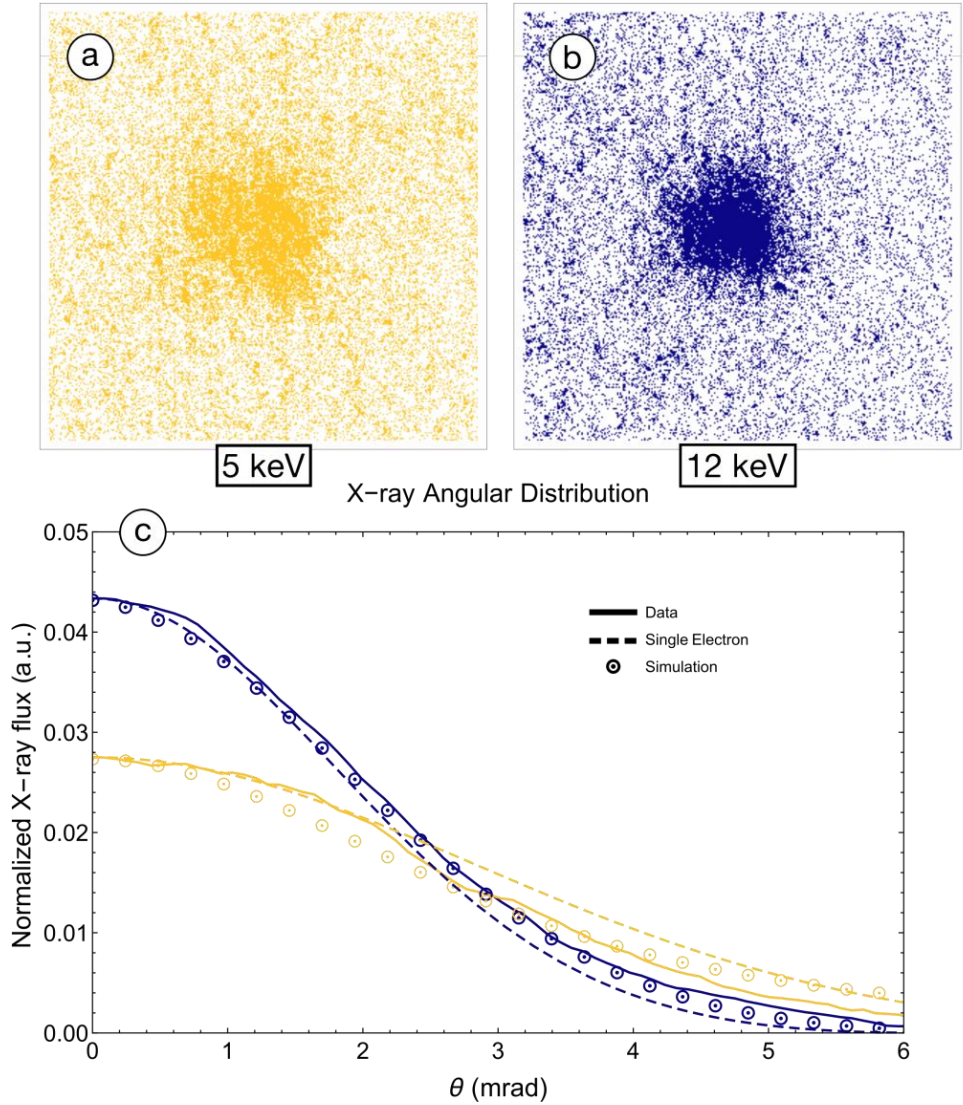

*Figure 5. The images in a) and b) are false-color raw images collected from the MCP. The angular distribution of X-rays has a characteristic opening angle that depends on the energy of the electrons involved in the Compton scattering process, $\theta \sim 1/\gamma$. Harder X-rays are produced by higher energy electrons and have a smaller opening angle, $3.4 \pm 0.1$ mrad. This is demonstrated in c), where the lineouts from a) and b) are compared to a Compton scattering simulation and the analytic expression for a single electron emission. The lineouts are normalized to their area.*

## Discussion

The results presented here show the successful merging of an IFEL accelerator with an ICS interaction point, to yield a first demonstration of a new class of all-optical X-ray light source. This source, owing to its unique dynamics properties, operated stably with highly reproducible shot-to-shot electron beam and X-ray characteristics. The maximum acceleration gradient achieved in the IFEL of $55$ MeV/m, obtained at laser intensities near $10^{13}$ W/cm$^2$, already exceeds that utilized in nearly all RF linear accelerators. Furthermore, the use of an optical pre-buncher enabled the fraction of accelerated electrons to be ~40%. These accelerated electrons were used to create $11.6 \text{ keV}$ X-rays with a photon yield of $1.22(\pm 0.16) \times 10^6$ per shot. It should be noted that not only is the method of creating these X-ray pulses unique, but they also have novel, heretofore unrealized properties owing to the temporal structure of the accelerated beam that produces the X-rays. The IFEL produces a distinctive micro-bunched electron beam, which has density spikes at the wavelength of the driving laser, in this case $CO_2$ at $10.3 \text{ μm}$. The temporal structure of the X-ray spectrum closely mimics that of the electron beam and in this way forms a train ($33 \text{ fs}$ period) bursts of photons. This periodic burst pattern is also produced in HHG sources[28] in the soft-X-ray spectral region, and has found compelling use in the stroboscopic probing of periodically driven (e.g. by a pulse of the source laser itself) physical systems[29]. The ICS source extends the spectral range such methods to hard X-rays and beyond.

The proof-of-principle experiment executed here establishes the IFEL acceleration scheme on solid footing as a driver for a 5th-generation, short wavelength light-source. It has been shown that the IFEL may indeed be applied as a compact, moderate-energy, optical accelerator that may serve as a driver for a flexible, laser-electron interaction-based source that accesses a highly useful range of photon energies, extending from $0.01$ to $10 \text{ MeV}$. As such, the results of this experiment are a promising step forward in advanced accelerator and light-source research, showing a novel approach that may be exploited to produce a short wavelength photon source having a number of desirable characteristics, including compactness, the potential to reach very high fluxes through recirculation, and the capability of producing femtosecond, periodic, stroboscopic X-rays.

## Acknowledgements
This work supported by the US DOE Office of High Energy Physics through contract DE-SC0009914, and US Dept. of Homeland Security Grant 2014-DN-077-ARI084-01, and the US DOE Office of Science SCGSR Graduate Student Research Fellowship program.

## Author Contributions
All authors contributed significantly to the writing of this paper.

## Author Information

Reprints and permissions information are available at www.nature.com/reprints. The authors declare no competing financial interests. Readers are welcome to comment on the online version of the paper. Correspondence and requests for materials should be addressed to I.I. Gadjev, gadjev@physics.ucla.edu.

# Methods

## Two pulse amplification

The staging of the IFEL accelerator and the Compton scattering IP requires two high-power $CO_2$ laser pulses separated by $500\ ps$, both of which are obtained from the terawatt $CO_2$ laser system at the BNL ATF. In order to create two laser pulses, the seeding pulse to the main amplifier was split prior to amplification. This was done with a Michelson interferometer type of beam-splitter, which output 2 co-propagating pulses – each with 25% of the original energy. These two seed pulses were individually amplified by a single discharge of the main amplifier. The energy extracted by each pulse from the main amplifier scales linearly with the energy of the seeding pulse. The output laser energy per pulse was stable with an average energy of $1.02 \pm 0.12\ J$. The IFEL acceleration gradient and the capture rate have a direct dependence on the magnitude of the electric field of the driving laser, as discussed in Duris, *et al.*[18] The tapering of the undulator is governed by the practical requirements on the capture rate and the acceleration gradient for a given amount of laser energy. At $1\ J$ laser energy per pulse, the undulator tapering was tuned to obtain 40% nominal capture rate at an acceleration gradient of 55 MeV/m.

## Spatial Overlap

Maintaining spatial overlap between the electron bunch and the $CO_2$ laser is imperative for sustaining the IFEL acceleration and for maximizing the ICS X-ray flux. To ensure proper overlap over the length of the undulator, the individual magnets of the undulator were tuned so that the trajectory of on-axis electrons is completely within the FWHM of the laser for the entire IFEL interaction. The tuning process relies on Hall-probe scans of the on-axis undulator field, which are then compared to a 3D magnetostatic simulations. It is important to keep the angle of exiting electrons close to zero, in order to transport the beam downstream. The electrons' exit trajectory is heavily influenced by both the final undulator magnets and the alignment of the undulator with respect to the propagation axis of the electron beam. An alignment He-Ne laser was used to define the electron beam axis and the undulator was aligned to this laser. To verify the transverse alignment of the undulator, beam position monitors (BPM) that reveal the transverse position of the He-Ne and e-beam were installed at the entrance and exit of the undulator.

## Final focusing for ICS

The number of scattered X-ray photons is proportional to the densities of the colliding electron bunch and $CO_2$ laser pulse. Therefore, a tight focus at the collision point would increase the flux of x-rays produced for a single collision event. On the other hand, the IFEL requires a significant laser electric field. Because the two $CO_2$ pulses have the same path and pass through the same optics, the optics must be arranged in such a way that the laser comes to a waist at two points in its path. The first focus is at the IFEL undulator. A NaCl lens with focal-length 4.5 m was used to achieve a $0.91\ mm$ waist at the IFEL undulator. The leading $CO_2$ pulse is refocused and reflected back onto the electron bunch by a Cu spherical mirror with a *f/#*=1.5 focal number. Depending on the divergence of the $CO_2$ at the mirror, we estimate that the focus at the ICS IP has a waist of $100 - 150\ \mu$m.

In order to focus the electron beam down to a comparable size, $1.8~\text{m}$ downstream of the undulator, a quadrupole triplet was employed. The transport from the end of the IFEL undulator through the focusing triplet and into the ICS IP can be written in terms of the initial beam $\sigma$-matrix and an effective transport matrix for the section, $\overline{\sigma_2} = M_{1\to 2}\,\overline{\sigma_1}\,M_{1\to 2}^{\dagger}$. The transport matrix for the quadrupole -triplet depends on the quadrupole strength parameter, $K_q = \frac{e\partial_x B_y}{\gamma\beta m_e c}$, which in turn is proportional to the magnetic field gradient and momentum of the electron. Therefore, if a beam with a certain emittance and energy is brought to a waist some distance away from the triplet, then a beam with the same emittance, but higher in energy will be focused at the same distance, if the quadrupole strengths are scaled by the ratio of the energies. In our case, we were able to use the quad-triplet to focus the $52~\text{MeV}$ beam to a transverse size of $\sigma_x \times \sigma_y = 140 \times 170~(\pm 10)~\mu m$ and then scale the quadrupole strengths by a factor of $82/52$, for the accelerated electrons. The achromatic effect of the final focus clearly observed in the spectrometer images in Figure 3, where the in-focus portion of the electron distribution is of higher energies. The scaled quadrupole currents brought the higher energy electrons to a focus at the IP, while lower energy electrons were no longer at a waist at the IP. This helped to increase the scattered flux of X-rays from higher energy electrons and decrease that of lower energy electrons. The ability to tune the final focus of the electron beam can in this way be used to suppress the production of lower energy X-rays.

In parts a), c), and d) of Figure 4, a start-to-end simulation of the pre-bunching, IFEL acceleration, final focus, and Compton scattering is presented. Part a) shows the electron beam transport and transverse focus into the ICS IP for the unaccelerated (yellow) and accelerated (blue) components of the electron beams. Since the final focusing quadrupole triplet is optimized for the accelerated beam, the low energy electrons do not contribute to the ICS X-ray spectrum, even though they make up a significant portion of the electron beam, as shown in Figure 4. c) and d). It is possible to focus the accelerated electrons into the ICS IP, because of the IFEL's unique ability to produce beams with a well-defined transverse emittance. In part b) of Figure 4., quadrupole scan measurements made after the IFEL undulator confirm that electron beam's normalized emittance is well-preserved during the IFEL acceleration. The normalized transverse emittance of the unaccelerated beam was measured to be $\varepsilon_n = 2.3 \pm 0.1~\mu m$, while that of the accelerated beam was measured to be $\varepsilon_n = 2.4 \pm 0.15~\mu m$. By properly matching the incoming electron beam $\beta$-function to the natural focusing of the IFEL undulator, we demonstrated for the first time conservation of the incoming beam emittance in the IFEL acceleration scheme.

### X-ray detection and filtering

The micro-channel plate (MCP) detector used to image the ICS X-rays has a KBr photocathode that converts X-ray photons to free electrons at a $50\%$ quantum efficiency[30–32]. The active area of the MCP has a radius of $20~\text{mm}$. Placed $76~\text{cm}$ after the ICS interaction point, makes its angle of acceptance, $\theta_{MCP} = 26.3~\text{mrad} > \theta_{X-\text{ray}} = \frac{1}{\gamma} = 6.25~\text{mrad}$. In our experimental conditions, the energy distribution of the Compton scattered X-rays follows that of the electron beam. It is thus possible to infer the X-ray spectrum based on the electron spectrometer data coupled with a knowledge of the electron beam focus. Figure 4. d) shows the simulation reconstructed X-ray spectrum corresponding to a typical IFEL electron beam. In practice, to estimate the flux of higher energy ICS X-rays, an Al foil was used to attenuate photons with energies below $6~\text{keV}$.